\documentclass[twocolumn,prl,showpacs]{revtex4}%
\usepackage{graphicx}%
\usepackage{amsmath}%
\usepackage{amsfonts}%
\usepackage{amssymb}

\def\d{{\partial}}
\def\s{{\sigma}}
\def\e{{\epsilon}}
\def\k{{ {\bf k} }}

\def\w{{\omega}}

\def\g{{\gamma}}

\def\r{{ {\bf r} }}

\begin{document}
\title{
Giant Orbital Hall Effect in Transition Metals: \\
Origin of Large Spin and Anomalous Hall Effects 
}
\author{H. \textsc{Kontani}$^{1}$, T. \textsc{Tanaka}$^{1}$, 
D.S. \textsc{Hirashima}$^{1}$, K. \textsc{Yamada}$^{2}$ 
and J. \textsc{Inoue}$^{3}$ }
\date{\today }

\begin{abstract}
In transition metals and their compounds, the orbital degrees of freedom 
gives rise to an orbital current, in addition to the ordinary spin and 
charge currents. 
We reveal that considerably large spin and anomalous Hall
effects (SHE and AHE) observed in transition metals 
originate from an orbital Hall effect (OHE). 
To elucidate the origin of these novel Hall effects, 
a simple periodic $s$-$d$ hybridization model is proposed as a generic model. 
The giant positive OHE originates
from the orbital Aharonov-Bohm phase factor, 
and induces spin Hall conductivity that is proportional to the 
spin-orbit polarization at the Fermi level, which is positive (negative) 
in metals with more than (less than) half-filling. 
\end{abstract}

\pacs{72.25.Ba,72.10.Bg,72.80.Ga}


\address{
$^1$Department of Physics, Nagoya University,
Furo-cho, Nagoya 464-8602, Japan. \\
$^2$Ritsumeikan University,
1-1-1 Noji Higashi, Kusatsu, Shiga 525-8577, Japan. \\
$^3$Department of Applied Physics, Nagoya University,
Furo-cho, Nagoya 464-8602, Japan.
}

\sloppy

\maketitle

The Hall effect, first discovered at the end of 19th century, has revealed the profound
nature of electron transport in metals and semiconductors via the
anomalous Hall effect (AHE) and (fractional) quantum Hall effects. 
It has recently been recognized that conventional semiconductors and metals
exhibit a spin Hall effect (SHE), which is the phenomenon where an electric
field induces a spin current (a flow of spin angular momentum $s_{z}$) in a
transverse direction \cite{Murakami-SHE,Sinova-SHE,Valenzuela,Saitoh,Kimura}.
Recently, a theory of the intrinsic Hall effect proposed by Karplus and Luttinger
 \cite{KL}, which occurs in multiband systems and
is independent of impurity scattering, has been intensively developed 
 \cite{Inoue-SHE,Kontani-Ru}.
In particular, a quantum SHE has also been predicted and experimentally 
confirmed \cite{Kane,SCZhang}.

The spin Hall conductivity (SHC) 
observed in transition metals has given rise to
further issues regarding the origin of the SHE, 
since the SHC observed in Pt exceeds 
$200\ \hbar e^{-1}\cdot \Omega ^{-1}\mathrm{cm}^{-1}$, which is 
approximately $10^{4}$ times larger than that of n-type semiconductors 
\cite{Kimura}, and the SHCs in Nb and Mo are negative \cite{Otani-Nb}. 
The large SHE and the sign change of the SHC in transition
metals has attracted much interest, and many theoretical studies of the SHE have
so far been conducted based on realistic multiband models for Ru-oxide 
\cite{Kontani-Ru} and various $4d$ and $5d$ metals \cite{Tanaka-4d5d}, 
including Au, W \cite{Yao}, and Pt \cite{Kontani-Pt,Guo-Pt}. 
The calculated results for
the SHC semi-quantitatively agree with the observed results.
The mechanism for the SHE has been explained in such a way that
spin-orbit interactions (SOI) and the phase of hopping integrals of electrons 
give rise to the Aharonov-Bohm (AB) effect, and therefore the conduction electrons 
are subject to an effective spin-dependent magnetic field.

\begin{figure}[!htb]
\includegraphics[width=.8\linewidth]{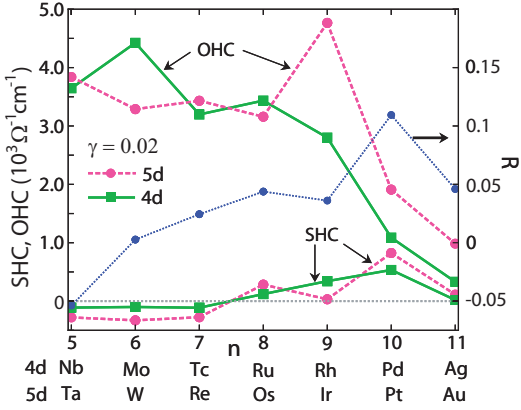}
\caption{\label{fig:SHE-OHE} 
(Color online)
SHC and OHC obtained in ref. \cite{Tanaka-4d5d}.
The body-centered cubic structure for $n=5,6$,
the hexagonal closed packed structure for $n=7,8$, and
the  face-centered cubic structure for $n=9\sim11$.
$n_d\approx n-1$ in each metal.
The SHC in Pt is approximately 1 in the unit of $e/2\pi a \approx 
10^3$ $\Omega^{-1}$cm$^{-1}$, where $a=0.39$ nm  is the lattice constant in Pt.
$R\equiv \langle l_zs_z\rangle_{\rm FS}$ is the spin-orbit polarization
for $5d$ metals.
}
\end{figure}

Since the transition metals have orbital degrees of 
freedom in addition to the spin and charge degrees of freedom, 
flow of the atomic orbital angular momentum ($l_z$),
that is, an orbital current,
may be realized in a nonequilibrium state.
In fact, several authors have predicted the emergence of a
large orbital Hall effect (OHE) \cite{p-OHE,Kontani-Ru,Tanaka-4d5d},
which is a phenomenon where an electric field induces a flow of 
$p$- and $d$-orbital angular momentum in a transverse direction.
In particular, the predicted orbital Hall conductivity (OHC) in
transition metals and oxides \cite{Kontani-Ru,Tanaka-4d5d}
is considerably larger than the SHC.
Figure \ref{fig:SHE-OHE} shows the OHC
and SHC calculated for transition metals using the Naval Research
Laboratory tight-binding (NRL-TB) model \cite{NRL1}. 
In each metal, the magnitude of the OHC exceeds the SHC, even 
in topological insulators 
(e.g., $\sim e/2\pi $ in graphene \cite{Kane} and HgTe \cite{SCZhang}).
Interestingly, while the SHC consistently changes its sign with 
the electron number $n$, as with several recent experiments 
\cite{Kimura,Otani-Nb}, the obtained OHC is almost independent of the SOI 
and is always \textit{positive}.
These prominent and universal features of orbital dynamics in metals,
independent of crystal and multiband structures,
have not been recognized until recently.

In spite of these remarkable features of the SHC and OHC
given in previous works \cite{Kontani-Ru,Tanaka-4d5d},
no physical origin of the ``giant OHE'' nor ``hidden relationship''
between OHE and SHE have been presented.
We will show below that the key phenomenon is the orbital Hall current,
which originates from the ``orbital AB phase factor'' 
that reflects the phase factor of the $d$-orbital wavefunction.
Then, the SHC is approximately given by the product of 
OHC and the spin-orbit polarization due to the SOI.
Even the AHE is understandable in the same concept.

In this Letter, we discuss these intrinsic Hall effects
in a unified way by proposing a 
simple $s$-$d$ hybridization model as a generic model, and explain why the
OHC is \textit{positive} and much larger than $e/2\pi $ in each transition
metal. 
We stress that
\textit{the large SHE in transition metals originates from the OHE} in
the presence of atomic SOI, not from the Dirac point monopole as in
semiconductors. The derived SHC is approximately proportional to the
spin-orbit polarization, which is positive (negative) in metals with more than
(less than) half-filling, which is consistent with recent experimental 
observations \cite{Saitoh,Kimura,Otani-Nb}. 
It is noted that the present 
OHE is different from the Hall effect of the angular momentum 
of $s$-electrons ${\bf r}\times {\bf p}$ discussed in Ref. \cite{OHE-Zhang}.

In ${\bar n}d$ transition metals (${\bar n}=3\sim5$ is the main quantum number), 
the electronic states near the Fermi level ($\mu $) are
constructed by ${\bar n}d$-orbitals, which hybridize
with free-electron-like $({\bar n}+1)s$- and $({\bar n}+1)p$-bands, 
and form a narrow band with a bandwidth in the order of $1$ eV \cite{NRL1}. 
The band structures of the transition
metals are well characterized by the lattice structures.
However, the OHC seems to be independent of the atomic species
of the transition metals as shown in Fig. \ref{fig:SHE-OHE}. 
The result suggests that the
details of the band structure may be irrelevant to the occurrence of the
large OHC in transition metals. 
Therefore, two-dimensional (2D) $s$-$d$ hybridization model with the 
orbital degree of freedom was adopted to study the
universal nature of the OHE in transition metals. The $s$-$d$ hybridization
model may also be the simplest version of the muffin-tin potential
approximation, where localized $d$-electrons can move only via $s$-$d$
hybridization.

The Hamiltonian for the 2D $s$-$d$ hybridization model is given by
$H_0= \sum_{\k,\s}{\hat a}_{\k\s}^\dagger {\hat H_0} {\hat a}_{\k\s}$, where
\begin{eqnarray}
{\hat H}_0 = \left(
\begin{array}[c]{ccc}
\e_\k & V_{\k L} & V_{\k -L} \\
V_{\k L}^* & E_d & 0 \\
V_{\k -L}^* & 0 & E_d
\end{array}
\right)  ,
 \label{eqn:Hsd}
\end{eqnarray}
and $^t{\hat a}_{\k\s}=(c_{\k\s},d_{\k L\s},d_{\k -L\s})$;
$c_{\k\sigma}$ and $d_{\k M\sigma}$ are annihilation operators for
the $s$- and $d$-electrons, respectively.
$\sigma=\pm1$ (or $\uparrow,\downarrow$) 
is the spin index, and $M=\pm L$ ($L$ is a positive integer)
represents the angular momentum of the $d$ electron.
$\epsilon_\k$ and $E_d$ represent the unhybridized
energies of the $s$- and $d$-electrons, respectively, 
and $V_{\k M}$ is the $s$-$d$ mixing potential.
In transition metals, 
the OHE and SHE are mainly caused by interorbit transitions
between $d_{yz}$ and $d_{zx}$ ($l_z=\pm1$) orbitals, and
$d_{xy}$ and $d_{x^2-y^2}$ ($l_z=\pm2$) orbitals \cite{Tanaka-4d5d,Kontani-Pt};
the former (latter) contributions can be obtained by letting $L=1$ ($L=2$) 
in the present 2D model.

To elucidate the universal properties of the OHC and SHC
that are independent of the detailed crystal structure,
we assume that $\epsilon_\k = \k^2/2m$ (plane wave).
As is well known, $e^{i\k\cdot\r}=\sum_n i^n J_n(kr)e^{in(\varphi_k-\varphi_r)}$,
where $J_n$ is the Bessel function,
$\varphi_k= \tan^{-1}(k_x/k_y)$ and $\varphi_r= \tan^{-1}(x/y)$.
Since the wavefunction of the $d$-electron is 
$\xi_M({\bf r})= f(r)e^{iM\varphi_r}$,
the $s$-$d$ mixing potential is given by
$V_{\k M} =\int (e^{i\k\cdot{\bf r}})^* H_0 \xi_M({\bf r})d{\bf r}
=V_0 e^{iM\varphi_k}$ in the extended Brillouin zone scheme 
\cite{Anderson,Hanzawa,Kontani94}.
(Note that $V_{\k M} \rightarrow V_{\k M}^\ast$ under the 
particle-hole transformation.)
Here, the $\k$-dependence of $V_0$ is neglected.
We will show that the phase factor of $V_{\k M}$ 
plays an essential role in the OHE, and the derived OHC takes a large value
irrespective of the nonconservation of $M=\pm L$.

\begin{figure}[!htb]
\includegraphics[width=.9\linewidth]{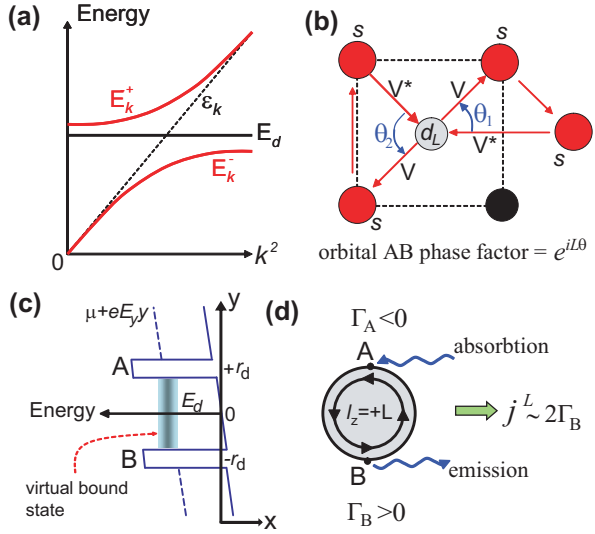}
\caption{\label{fig:sd-model} 
(Color online)
(a) Schematic band structure of the $s$-$d$ hybridization model 
in Eq. (\ref{eqn:Hsd}).
(b) Examples of the clockwise motions of electrons along the nearest 
three sites, which give the orbital AB phase factor
$e^{i L \theta_i}$ with $\theta_i>0$.
(c) Localized $d$-orbital state and the Fermi level of
conduction electrons under $E_y$.
(d) A semiclassical explanation for the Hall current
($\perp {\bf E}$) due to the angular momentum conservation.
An left-moving electron is converted to right-moving 
by $l_z=+L$ orbital.
}
\end{figure}

The present model is similar to the periodic Anderson model, which has been 
intensively studied as an effective model for $f$-electron systems 
\cite{Anderson,Hanzawa,Kontani94}.
The close similarity between the periodic Anderson model and the $d$-$p$ model 
has been indicated in the previous study of the AHE \cite{Miyazawa}.
As is well-known, the hybridization band of Eq. (\ref{eqn:Hsd}) is given as
$E_\k^\pm= \frac12 [(\e_\k+E_d)\pm\sqrt{(\e_\k-E_d)^2+2V_0^2}]$,
the bandstructure of which is shown in Fig. \ref{fig:sd-model}(a).
In the metallic state, the Fermi level $\mu$ is located in the
upper ($E_\k^+$) or lower ($E_\k^-$) branch, and the 
relationship $(\mu-\e_{k_{\rm F}})(\mu-E_d)=2V_0^2$ holds.
The relation $N_d(0)/N_s(0)= 2V_0^2/(\mu-E_d)^2 \gg 1$
is satisfied in transition metals,
where $N_s(0)=m/2\pi$ is the $s$-electron density of states (DOS) 
per spin in 2D, and $N_d(0)$ is the $d$-electron DOS at $\mu$.

According to the linear-response theory, the intrinsic Hall conductivity 
is given by the summation of the Fermi surface term ($I$-term) and 
the Fermi sea term ($I\!I$-term) \cite{Streda}.
In previous work, we have shown that the $I$-term is dominant 
in many metals \cite{Kontani06,Tanaka-4d5d,Kontani-Pt}.
Using the $3\times3$ Green function 
${\hat G}_\k^0(\w)=(\w+\mu-{\hat H}_0)^{-1}$,
the $I$-term of the OHC at $T=0$ is given by 
\cite{Kontani-Ru,Tanaka-4d5d}
\begin{eqnarray}
O_{xy}^z &=& \frac{1}{\pi N}\sum_\k {\rm Tr}\left[ 
{\hat J}_x^O {\hat G}_\k^0(i\gamma){\hat J}_y^C {\hat G}_\k^0(-i\gamma) 
\right] ,
 \label{eqn:OHE-formula}
\end{eqnarray}
where ${\hat J}_y^C$ is the $y$-component of the charge current,
which is given by $-e\d {\hat H}_0/\d k_y$.
In the present model,
\begin{eqnarray}
{\hat J}_y^C = -e\left(
\begin{array}[c]{ccc}
k_y/m & iL \frac{k_x}{k^2} V_{\k L} & -iL \frac{k_x}{k^2} V_{\k L}^* \\
-iL \frac{k_x}{k^2} V_{\k L}^* & 0 & 0 \\
 iL \frac{k_x}{k^2} V_{\k L} & 0 & 0
\end{array}
\right)  ,
 \label{eqn:JCy}
\end{eqnarray}
where $-e$ is the charge of the electron.
In addition, ${\hat J}_x^O$ in Eq. (\ref{eqn:OHE-formula}) is the $x$-component 
of the orbital current, which is given by 
${\hat J}_x^O =\{{\hat J}_x^C,{\hat l}_z \}/(-2e)$ \cite{Tanaka-4d5d}, 
where $({\hat l}_z)_{i,j}=L(\delta_{i,2}-\delta_{i,3})\delta_{i,j}$.
In the present model,
\begin{eqnarray}
{\hat J}_x^O = \frac12 iL^2 \frac{k_y}{k^2} \left(
\begin{array}[c]{ccc}
0 & V_{\k L} & V_{\k L}^* \\
-V_{\k L}^* & 0 & 0 \\
-V_{\k L} & 0 & 0
\end{array}
\right)  .
 \label{eqn:JOx}
\end{eqnarray}
Here, the relations
$\d V_{\k L}/\d k_x= -iL({k_y}/{k^2})V_{\k L}$ and
$\d V_{\k L}/\d k_y= iL({k_x}/{k^2})V_{\k L}$ are used \cite{Kontani94}.
Thus, the momentum derivative of the $s$-$d$ mixing potential
gives rise to an anomalous velocity that is perpendicular to $\k$:
The OHC is proportional to $\langle ({\hat J}_y^C)_{1,1} 
({\hat J}_x^O)_{1,2}V_{\k L}^*\rangle_{\rm FS} \ne 0$.

By inserting Eqs. (\ref{eqn:JCy}) and (\ref{eqn:JOx})
into Eq. (\ref{eqn:OHE-formula}), the OHC in the present model is 
simply obtained as
\begin{eqnarray}
O_{xy}^z 
&=& \frac{eL^2}{4\pi}\cdot\frac{2V_0^2}{m N} \sum_\k
  \frac{\gamma}{|g_\k(i\gamma)|^2},
 \label{eqn:SOI}
\end{eqnarray}
where $N$ is the number of $\k$-points,
$g_\k(\w)=(\w+\mu-E_\k^+)(\w+\mu-E_\k^-)$, 
and $\g$ is the 
damping rate (due to impurity potentials).
The obtained OHC is finite even if SOI is absent
 \cite{Kontani-Ru,Tanaka-4d5d,Kontani-Pt}.
Since $\lim_{\g\rightarrow+0} \g/(x^2+\g^2)=\pi\delta(x)$
and $\frac1N \sum_k\delta(\mu-E_\k^\pm)=N_d(0)$,
We obtain that
%
\begin{eqnarray}
O_{xy}^z &=&
 \frac{eL^2}{4m} \left[ N_d(0)\frac{2V_0^2}{(E_\k^+-E_\k^-)^2}\right]
\sim \frac{e L^2}{4\pi},
 \label{eqn:SOI-ap}
\end{eqnarray}
where we used that the term in the bracket is $\sim N_s(0)$
when $N_d(0)/N_s(0)\gg1$.
Thus, the OHC takes an approximate universal positive value,
independent of the model parameters.
For the 3D $s$-$d$ hybridization model,
$V_{\k M}$ is proportional to the spherical harmonics with $L=2$.
In this model, we obtain $O_{xy}^z \sim (e/2\pi a)\cdot \pi$ 
if we put $k_{\rm F}\sim\pi/a$, where $a$ is the lattice spacing.
This result is qualitatively consistent with 
the results in Fig. \ref{fig:SHE-OHE} for $n\le9$.
(OHCs in Au and Ag are small since $n_d=10$.)
We have verified that the $I\!I$-term is negligibly small.

The reason why the giant OHE emerges
in transition metals based on a 2D tight-binding model
with three orbitals ($s$ and $d_{\pm L}$) at each site is now discussed.
Long-range hopping integrals must be considered to reproduce 
a free-electron-like $s$-electron dispersion \cite{NRL1}.
Figure \ref{fig:sd-model}(b) illustrates two examples of the clockwise 
motion of electrons along the nearest three sites
[$d_L\rightarrow s \rightarrow s \rightarrow d_L$];
as discussed in Ref. \cite{Tanaka-4d5d},
these are important processes for the OHE,
because the $s$-$d$ hopping integrals are much larger than 
the $d$-$d$ hopping integrals.
Therein, the electron acquires the phase factor $e^{iL\theta}$
due to the angular dependence of the mixing potential in real-space, 
$V_{L}(\r)\propto e^{iL\varphi_r}$,
where $\theta$ is the angle between the incoming and outgoing electron
($\theta=\pi/4$ or $\pi/2$ in this figure). 
This can be interpreted as the ``orbital AB phase'' 
given by the effective magnetic flux $\phi_0\cdot(L\theta/2\pi)$ through the
area of the triangle, where $\phi_0=2\pi\hbar/e$ is the flux quantum.
It is simple to check whether any of the other three-site clockwise motions 
cause the factor $e^{iL\theta}$ with $\theta>0$.
Therefore, the $d$-electron with $l_z=\pm L$ is subject to 
the huge and {\it positive} 
effective magnetic field that reaches $\sim \pm \phi_0/a^2$.
This is the origin of the giant positive OHE in the order of $e/2\pi$
in transition metals.

We also present a simple semiclassical explanation, in that the OHE 
is a natural consequence of the anisotropy in $s$-$d$ hybridization
under the electric field $E_y$.
In the muffin-tin model,
the localized $d$-electron (virtual bound state) 
moves to the conduction band via the $s$-$d$ mixing potential \cite{Hewson}
as shown in Figs. \ref{fig:sd-model} (c) and (d).
According to the Fermi's golden rule,
the $d\!\rightarrow\! s$ tunneling probability at point B is given by
$\Gamma_B \sim \int_{\mu_B}^\mu d\w N_d(\w)|V_0|^2N_s(\w)$,
where $\mu_B=\mu-eE_y r_d$ is the electrochemical potential at B under $E_y$
Then, $\Gamma_A =-\Gamma_B$ and $\Gamma_B \sim e E_y r_d$ 
since $N_s(0)\sim|\mu-\e_{k_{\rm F}}|^{-1}$ and $N_d(0)\sim|\mu-E_d|^{-1}$.
Because of the angular momentum conservation,
the velocity of emitted (absorbed) electron at point B (A) has $x$-component;
$v_x^L \sim (-)L/r_d m$.
If we assume that the tunneling $s$-electron hybridizes to 
one of the neighboring sites and $l_z=\pm L$ is quasi-conserved,
the current of the successive tunneling electron will be
$j_x^{\pm L} \sim \pm n |\Gamma_{d\rightarrow s}| 
\cdot a \sin \varphi_o$,
where $n$ is the electron density  
and $\sin \varphi_o \equiv |v_x^L|/v_F \sim O(1)$.
Therefore, the estimated transverse $d$-orbital current density is 
\begin{eqnarray}
j_x^O &\sim& \sum_{M=\pm L} M j_x^{M} 
\sim eL^2E_y n a/ m v_{\rm F}.
 \label{eqn:OHC}
\end{eqnarray}
Since $a \sim \pi k_{\rm F}^{-1}$ and $n \sim a^{-2}$,
$O_{xy}^z\equiv j_x^O/E_y$ is in the order of $+eL^2$.
Thus, Eq. (\ref{eqn:SOI-ap}) is reproduced (aside from a numerical factor)
by this semiclassical consideration.

We note that the partial wave of the $l_z=L$ channel,
$\psi_L(\r)\propto J_L(kr)e^{iL\varphi_r}$,
has a small overlap integral between the nearest sites, 
$\int \psi_L^*(\r) \psi_{L'}(\r+a{\hat x}) d\r$, for $L'=-L$, 
due to the phase factor in $\psi_L(\r)$.
For this reason, $l_z$ is quasi-conserved
when the tunneling $s$-electron hybridizes to the $d$-orbital at a
neighboring site.
Mathematically, the anomalous velocity is given by taking the 
gradient of the phase factor in $\psi_L(\r)$; see eq. (\ref{eqn:JOx}).

Next, we discuss the SHE in the presence of the atomic SOI; 
$\lambda \sum_i{\bf l}_i\cdot{\bf s}_i$ ($\lambda>0$).
Since $\langle M|l_\nu|M' \rangle=0$ for $\nu=x,y$ in the present 2D model,
the atomic SOI for the $\s$-spin is given by
$({\hat H}_{\lambda}^\s)_{i,j}
 = (\lambda\s/2)L(\delta_{i,2}-\delta_{i,3})\delta_{i,j}$.
Also, only the $z$-component of SOI is significant
for the SHE in real transition metals \cite{Tanaka-4d5d,Kontani-Pt}.
Using the Green function 
${\hat G}_{\k\s}(\w)=(\w+\mu-{\hat H}_0-{\hat H}_{\lambda}^\s)^{-1}$,
the SHC is given by
\begin{eqnarray}
\sigma_{xy}^z &=& \frac{1}{2\pi N}\sum_{\k,\sigma} \frac{\sigma}{-2e}
{\rm Tr}\left[ {\hat J}_x^C {\hat G}_{\k\sigma}(i\gamma)
{\hat J}_y^C {\hat G}_{\k\sigma}(-i\gamma) \right] .
 \label{eqn:SI}
\end{eqnarray}
If the $\lambda$-dependence of eigenenergies is neglected,
which corrects the SHC of order $O(\lambda^3)$ to,
Eq. (\ref{eqn:SI}) becomes
\begin{eqnarray}
\sigma_{xy}^z &\approx& 2R/L^2 \cdot O_{xy}^z,
 \label{eqn:SI-ap}
\end{eqnarray}
where $R\equiv \langle{\hat l}_z{\hat s}_z\rangle_{\rm FS}$
represents the spin-orbit polarization ratio
due to the SOI at the Fermi level, which is given by
$R=L\lambda/(\mu-E_d)$ in the present model up to $O(\lambda)$. 
Thus, the SHC is positive (negative) when $\mu$ is located
in the upper branch $E_\k^+$ (lower branch $E_\k^-$).

It is natural to expect that the relationship in Eq. (\ref{eqn:SI-ap}) holds
in real transition metals, 
where the spin-orbit polarization ratio is defined as
$R=\sum_m \int_{\rm FS}\langle{\hat l}_z{\hat s}_z\rangle_{\k,m} dS_{\k,m}
/\sum_m \int_{\rm FS} dS_{\k,m}$ in real systems, where $m$ is the band index.
To verify this expectation, $R$ is shown for $5d$ metals given by the 
NRL-TB model in Fig. \ref{fig:SHE-OHE}:
The obtained $R$ is positive (negative) in metals with more than 
(less than) half-filling, which is consistent with Hund's rule.
The qualitative similarity between the SHC and $R$ in Fig. \ref{fig:SHE-OHE}
indicates that the spin current ${\vec j}^S$ is induced by the 
orbital current ${\vec j}^O$ in proportion to $R$, and therefore
the relationship in Eq. (\ref{eqn:SI-ap}) holds approximately 
for various metals.
(In fact, $\lambda l_zs_z$ provides the dominant contribution 
to the SHE \cite{Tanaka-4d5d}.)
As a result, the present analysis based on a simple $s$-$d$ hybridization 
model captures the overall behavior of the OHE and SHE in transition metals.

In the low resistivity regime, the intrinsic SHC is given by integrating the 
$\k$-space Berry curvature of Bloch wavefunction 
(Berry curvature term)
 \cite{Murakami-SHE,Sinova-SHE,Kontani-Pt,Guo-Pt,Tanaka-4d5d}.
In fact, previous studies based on the tight-binding models
 \cite{Kontani-Pt,Tanaka-4d5d} 
and the band calculation \cite{Guo-Pt}
had succeeded in reproducing experimental SHC's in several 
transition metals with low resistivity, 
both in magnitude and sign \cite{Kimura,Otani-Nb}.
The present study has shown that the large Berry curvature 
in transition metals, the origin of which had been unclear in previous studies,
originates from the $d$-orbital angular momentum.
We have revealed the existence of the real-space orbital Berry phase 
(=AB phase), which causes not only the giant positive OHE without 
using the SOI, but also the SHE and AHE if $R\ne0$.
By virtue of this scheme, 
hidden relationships between the OHE and SHE have been derived.
Although the OHE is indirectly observed via SHE and AHE,
it is interesting to detect the OHE directly:
We propose that a mesoscopic ``H-shape'' circuit will be useful,
which was originally used to measure the SHC in semimetals 
\cite{Hankiewicz}.

To summarize, we have revealed that the giant positive OHE in transition metals
originates from the orbital AB phase 
due to the $d$-orbital angular momentum, without necessity of 
any special bandstructure (e.g., Dirac point monopole at $\mu$).
We have shown that the OHE is the essential phenomenon,
and it induces the large SHE (AHE) in paramagnetic (ferromagnetic)
metals in the presence of SOI.
The sign of the SHC is
equal to that of the spin-orbit polarization (Hund's rule),
which is consistent with recent experimental observations
 \cite{Saitoh,Kimura,Otani-Nb}.
An intuitive explanation for the intrinsic Hall effect
in real-space is presented in Figs. \ref{fig:sd-model} (c) and (d).

We are grateful to 
E. Saitoh, T. Kimura and Y. Otani
for valuable comments and discussions.
This study has been supported by Grants-in-Aids for Scientific
Research from MEXT, Japan.



\begin{thebibliography}{99}

\bibitem{Murakami-SHE}
S. Murakami et al.,
Phys. Rev. B {\bf 69} (2004) 235206.

\bibitem{Sinova-SHE}
J. Sinova et al.,
Phys. Rev. Lett. {\bf 92} (2004) 126603.

\bibitem{Valenzuela}
S. O. Valenzuela and M. Tinkham: Nature {\bf 442} (2006) 176.

\bibitem{Saitoh} 
E. Saitoh et al.,
Appl. Phys. Lett. {\bf 88} (2006) 182509.

\bibitem{Kimura}
T. Kimura, Y. Otani, T. Sato, S. Takahashi, and S. Maekawa:
Phys. Rev. Lett. {\bf 98} (2007) 156601.

\bibitem{KL}
R. Karplus and J.M. Luttinger, Phys. Rev. {\bf 95}, 1154 (1954);
J. M. Luttinger, Phys. Rev. {\bf 112}, 739 (1958).

\bibitem {Inoue-SHE}
J. Inoue et al.,
Phys. Rev. B\textbf{70} (2004) 041303(R).

\bibitem{Kontani-Ru}
H. Kontani et al., 
Phys. Rev. Lett. {\bf 100}, 096601 (2008) .

\bibitem{Kane}
C.L. Kane and E.J. Mele, Phys. Rev. Lett. {\bf 95} (2005) 146802.

\bibitem{SCZhang} 
B.A. Bernevig et al., Science {\bf 314}, 1757 (2006).


\bibitem{Otani-Nb}
The observed SHC in Nb is 
$-16\ \hbar e^{-1}\cdot \Omega ^{-1}\mathrm{cm}^{-1}$;
Y. Otani et al, (unpublished).

\bibitem{Tanaka-4d5d}
T. Tanaka et al.,
Phys. Rev. B {\bf 77}, 165117 (2008).

\bibitem{Yao}
Y. Yao and Z. Fang , Phys. Rev. Lett. {\bf 95}, 156601 (2005);
SHCs in W and Au obtained by them are much larger than those in Fig. 
\ref{fig:SHE-OHE}.

\bibitem{Kontani-Pt}
H. Kontani et al.,
J. Phys. Soc. Jpn. {\bf 76} (2007) 103702. 

\bibitem{Guo-Pt} 
G.Y. Guo et al., 
Phys. Rev. Lett. {\bf 100} (2008) 096401.

\bibitem{p-OHE} 
B.A. Bernevig et al, Phys. Rev. Lett. {\bf 95}, 066601 (2005).

\bibitem{NRL1} 
D. A. Papaconstantopoulos and M.J. Mehl: 
J. Phys.: Condens. Matter {\bf 15} (2003) R413.

\bibitem{OHE-Zhang}
S. Zhang and Z. Yang, Phys. Rev. Lett. \textbf{94}, 066602 (2005).

\bibitem{Anderson}
Z. Zou and P. W. Anderson, Phys. Rev. Lett. {\bf 57}, 2073 (1986).

\bibitem{Hanzawa}
K. Hanzawa et al.,
Prog. Theor. Phys. {\bf 81} (1989) 960.

\bibitem{Kontani94}
 H. Kontani and K. Yamada: J. Phys. Soc. Jpn. {\bf 63} (1994) 2627.

\bibitem{Miyazawa}
M. Miyazawa et al.,
J. Phys. Soc. Jpn. {\bf 68} (1999) 1625.

\bibitem{Streda}
P. Streda, J. Phys. C: Solid State Phys. {\bf 15}, L717 (1982).

\bibitem{Kontani06}
H. Kontani et al., 
Phys. Rev. B {\bf 75} (2007) 184416.

\bibitem{Hewson}
A.C. Hewson, {\it The Kondo Problem to Heavy Fermions} 
(Cambridge Univ. Press, Cambridge, 1993). 


\bibitem{Hankiewicz}
E.M. Hankiewicz et al., Phys. Rev. B {\bf 70}, 241301(R) (2004).

\end{thebibliography}
\end{document}